\documentclass[nopreprintline,3p,authoryear]{elsarticle}

\usepackage{stackengine}
\setstackEOL{\\}
\usepackage[pages=some,placement=bottom,color=black,opacity=1,scale=1,vshift=20pt]{background}

\usepackage[utf8]{inputenc}

\usepackage{booktabs}

\usepackage[mode=image]{standalone}

\usepackage[font=footnotesize]{subcaption}

\usepackage{microtype}

\usepackage[bottom]{footmisc}

\usepackage[nopostdot,nonumberlist]{glossaries}
\setacronymstyle{long-short}  
\newacronym{ae}{AE}{autoencoder}
\newacronym{cae}{cAE}{correspondence autoencoder}
\newacronym{mfccs}{MFCCs}{Mel-frequency cepstral coefficients}
\newacronym{plps}{PLPs}{perceptual linear prediction coefficients}
\newacronym{dtw}{DTW}{dynamic time warping}
\newacronym{asr}{ASR}{automatic speech recognition}
\newacronym{utd}{UTD}{unsupervised term discovery}
\newacronym{ap}{AP}{average precision}
\newacronym{sw}{SW}{same-word}
\newacronym{dw}{DW}{different-word}
\newacronym{swsp}{SWSP}{same-word same-speaker}
\newacronym{swdp}{SWDP}{same-word different-speaker}
\newacronym{dwsp}{DWSP}{different-word, same-speaker}
\newacronym{dwdp}{DWDP}{different-word, different-speaker}
\newacronym{gmm}{GMM}{Gaussian mixture model}
\newacronym{hmm}{HMM}{hidden Markov model}
\newacronym{ubm}{UBM}{universal background model}
\newacronym{dnn}{DNN}{deep neural network}
\newacronym{rms}{RMS}{root mean square}
\newacronym{vtln}{VTLN}{vocal tract length normalization}
\newacronym{lvtln}{LVTLN}{linear VTLN}
\newacronym{zrsc}{ZRSC}{Zero Resource Speech Challenge}
\newacronym{bnfs}{BNFs}{bottleneck features}
\newacronym{cmn}{CMN}{cepstral mean normalization}
\newacronym{cvn}{CVN}{cepstral variance normalization}
\newacronym{cmvn}{CMVN}{cepstral mean and variance normalization}
\newacronym{sat}{SAT}{speaker adaptive training}
\newacronym{wsj}{WSJ}{Wall Street Journal}
\newacronym{dpgmm}{DPGMM}{Dirichlet Process Gaussian mixture model}
\newacronym{tdnn}{TDNN}{time-delay neural network}
\newacronym{sgmm}{SGMM}{subspace GMM}
\newacronym{fdlp}{FDLP}{frequency-domain linear prediction}

\newacronym{fMLLR}{fMLLR}{feature-space maximum likelihood linear regression}
\newacronym{LHUC}{LHUC}{learning hidden unit contributions}  

\journal{Computer Speech and Language}

\begin{document}
\BgThispage
\backgroundsetup{contents={\footnotesize%
    \Longstack{\textit{© 2020. Accepted for publication in Computer Speech and Language.}\\
      \textit{This manuscript version is made available under the CC-BY-NC-ND 4.0 license:}\\
      \textit{http://creativecommons.org/licenses/by-nc-nd/4.0/}}}}

\begin{frontmatter}
\title{Multilingual and Unsupervised Subword Modeling\\ for Zero-Resource Languages}
\author[1]{Enno Hermann\corref{cor}}
\ead{enno.hermann@idiap.ch}

\author[2]{Herman Kamper}
\ead{kamperh@sun.ac.za}

\author[1]{Sharon Goldwater}
\ead{sgwater@inf.ed.ac.uk}

\cortext[cor]{Corresponding author. Present address: Idiap Research Institute, Rue Marconi 19, Martigny 1920, Switzerland.}
\address[1]{School of Informatics, University of Edinburgh, Edinburgh EH8 9AB, U.K.}
\address[2]{Department of E\&E Engineering, Stellenbosch University, Stellenbosch 7600, South Africa}

\begin{abstract}
  Subword modeling for zero-resource languages aims to learn low-level
  representations of speech audio without using transcriptions or other
  resources from the target language (such as text corpora or pronunciation
  dictionaries). A good representation should capture phonetic content and
  abstract away from other types of variability, such as speaker differences and
  channel noise. Previous work in this area has primarily focused unsupervised
  learning from target language data only, and has been evaluated only
  intrinsically. Here we directly compare multiple methods, including some that
  use only target language speech data and some that use transcribed speech from
  other (non-target) languages, and we evaluate using two intrinsic measures as
  well as on a downstream unsupervised word segmentation and clustering task. We
  find that combining two existing target-language-only methods yields better
  features than either method alone. Nevertheless, even better results are
  obtained by extracting target language bottleneck features using a model
  trained on other languages. Cross-lingual training using just one other
  language is enough to provide this benefit, but multilingual training helps
  even more. In addition to these results, which hold across both intrinsic
  measures and the extrinsic task, we discuss the qualitative differences
  between the different types of learned features.
\end{abstract}

\end{frontmatter}

\glsresetall 

\section{Introduction}
Recent years have seen increasing interest in 
speech technology for ``zero-resource'' languages, where systems must be developed for a target language without using
transcribed data or other hand-curated resources from that language. Such
systems could potentially be applied to tasks such as endangered language
documentation or query-by-example search for languages without a written form.
One challenge for these systems, highlighted by the \gls{zrsc} shared tasks of
2015 \citep{Versteegh2015} and 2017 \citep{Dunbar2017}, is to improve subword
modeling, i.e., to extract or learn speech features from the target language
audio. Good features should be more effective at discriminating between
linguistic units, e.g. words or subwords, while abstracting away from factors
such as speaker identity and channel noise.

The ZRSCs were motivated largely by questions in artificial intelligence and
human perceptual learning, and focused on approaches where no transcribed data
from {\em any} language is used. Yet from an engineering perspective it also
makes sense to explore how training data from higher-resource languages can be
used to improve speech features in a zero-resource language.

This paper explores several methods for improving subword modeling in
zero-resource languages, either with or without the use of labeled data from
other languages. Although the individual methods are not new, our work provides
a much more thorough empirical evaluation of these methods compared to the
existing literature. We experiment with each method both alone and in
combinations not tried before, and provide results across a range of target
languages, evaluation measures, and tasks.

We start by evaluating two methods for feature extraction that are trained using
(untranscribed) target language data only: traditional \gls{vtln} and the
\gls{cae} proposed more recently by \citet{Kamper2015}. The \gls{cae} learns to
abstract away from signal noise and variability by training on pairs of speech
segments extracted using an \gls{utd} system---i.e., pairs that are likely to be
instances of the same word or phrase. We confirm previous work showing that
\gls{cae} features outperform \gls{mfccs} on a word discriminability task,
although we also show that this benefit is not consistently better than that of
simply applying \gls{vtln}. More interestingly, however, we find that applying
\gls{vtln} to the input of the \gls{cae} system improves the learned features
considerably, leading to better performance than either method alone. These
improvements indicate that \gls{cae} and \gls{vtln} abstract over different
aspects of the signal, and suggest that \gls{vtln} might also be a useful
preprocessing step in other recent neural-network-based unsupervised
feature-learning methods.

Next, we explore how multilingual annotated data can be used to improve feature
extraction for a zero-resource target language. We train multilingual \gls{bnfs}
on between one and ten languages from the GlobalPhone collection and evaluate on
six other languages (simulating different zero-resource targets). We show that
training on more languages consistently improves performance on word
discrimination, and that the improvement is not simply due to more training
data: an equivalent amount of data from one language fails to give the same
benefit. In fact, we observe the largest gain in performance when adding the
second training language, which is already better than adding three times as
much data from the same language. Moreover, when compared to our best results
from training unsupervised on target language data only, we find that \gls{bnfs}
trained on just a single other language already outperform the
target-language-only training, with multilingual \gls{bnfs} doing better by a
wide margin.

Although multilingual training outperforms unsupervised target-language
training, it could still be possible to improve on the multilingual \gls{bnfs}
by using them as inputs for further target-language training. To test this
hypothesis, we passed the multilingual \gls{bnfs} as input to the
\gls{cae}. When trained with \gls{utd} word pairs, we found no benefit to this
method. However, training with manually labeled word pairs did yield
benefits, suggesting that this type of supervision can help improve on the
\gls{bnfs} if the word pairs are sufficiently high-quality.

The results above were presented as part of an earlier conference version of
this paper \citep{Hermann2018}. Here, we expand upon that work in several ways.
First, we include new results on the corpora and evaluation measures used in the
\gls{zrsc}, to allow more direct comparisons with other work. In doing so, we
also provide the first set of results on identical systems evaluated using both
the same-different and ABX evaluation measures. This permits the two measures
themselves to be better compared. Finally, we provide both a qualitative
analysis of the differences between the different features we extract, and a
quantitative evaluation on the downstream target-language task of unsupervised
full-coverage speech segmentation and clustering using the system of
\citet{Kamper2017}. This is the first time that multilingual features are used in
such a system, which performs a complete segmentation of input speech into
hypothesized words. As in our intrinsic evaluations, we find that the
multilingual \gls{bnfs} consistently outperform the best unsupervised \gls{cae}
features, which in turn outperform or do similarly to MFCCs.

\section{Unsupervised Training, Target Language Only}
\label{sec:unsupervised}

We start by investigating how unlabeled data from the target language alone can
be used for unsupervised subword modeling and how speaker normalisation with
\gls{vtln} can improve such systems. Below we first review related work and
provide a brief introduction to the \gls{cae} and \gls{vtln} methods. We then
describe our experiments directly comparing these methods, both alone and in
combination.

\subsection{Background and Motivation}
\label{sec:cae}

Various approaches have been applied to the problem of unsupervised subword
modeling. Some methods work in a strictly bottom-up fashion, for example by
extracting posteriorgrams from a (finite or infinite) Gaussian mixture model
trained on the unlabeled data~\citep{Zhang2009,Huijbregts2011,Chen2015}, or by
using neural networks to learn representations using
autoencoding~\citep{Zeiler2013,Badino2014,Badino2015} or other loss
functions~\citep{Synnaeve2016}. Other methods incorporate weak top-down
supervision by first extracting pairs of similar word- or phrase-like units
using unsupervised term detection, and using these to constrain the
representation learning. Examples include the \acrfull{cae}~\citep{Kamper2015}
and ABNet~\citep{Synnaeve2014}. Both aim to learn representations that make
similar pairs even more similar; the ABNet additionally tries to make different
pairs more different.

\begin{figure}
\centering
  \includegraphics[width=0.7\columnwidth]{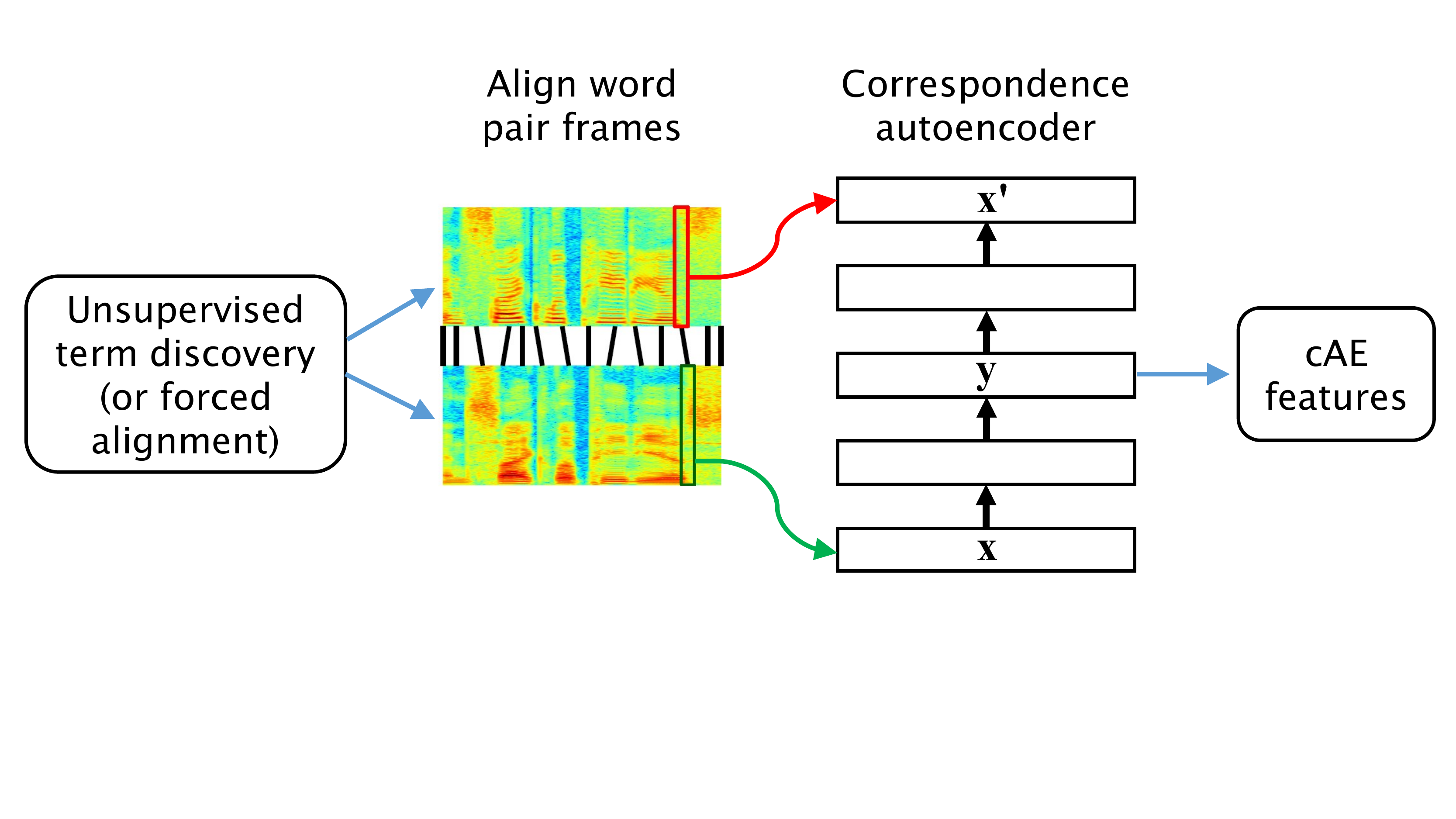}
  \vspace{-5em}
  \caption{Correspondence autoencoder training procedure (see section~\ref{sec:cae}).}
  \label{f:cae}
\end{figure}

In this work we use the \gls{cae} in our experiments on unsupervised
representation learning, since it performed well in the 2015 ZRSC, achieved some
of the best-reported results on the same-different task (which we also
consider), and has readily available code. As noted above, the \gls{cae}
attempts to normalize out non-linguistic factors such as speaker, channel,
gender, etc., by using top-down information from pairs of similar speech
segments. Extracting \gls{cae} features requires three steps, as illustrated in
Figure~\ref{f:cae}. First, an \acrfull{utd} system is applied to the target
language to extract pairs of speech segments that are likely to be instances of
the same word or phrase. Each pair is then aligned at the frame level using
\gls{dtw}, and pairs of aligned frames are presented as the input $\mathbf{x}$
and target output $\mathbf{x'}$ of a \gls{dnn}. After training, a middle layer
$\mathbf{y}$ is used as the learned feature representation.

The \gls{cae} and other unsupervised methods described above implicitly aim to
abstract away from speaker variability, and indeed they succeed to some extent
in doing so~\citep{Kamper2017}. Nevertheless, they provide less explicit speaker
adaptation than standard methods used in supervised \gls{asr}, such as
\gls{fMLLR}~\citep{Gales1998}, \gls{LHUC}~\citep{Swietojanski2016} or
i-vectors~\citep{Saon2013}. Explicit speaker adaptation seems to have attracted
little attention until recently~\citep{Zeghidour2016,Heck2016,Tsuchiya2018} in
the zero-resource community, perhaps because most of the standard methods assume
transcribed data is available.

Nevertheless, recent work suggests that at least some of these methods may be
applied effectively even in an unsupervised setting. In particular, 
\citet{Heck2017,Heck2018} won the \gls{zrsc} 2017 using a typical \gls{asr}
pipeline with speaker adaptive fMLLR and other feature transforms. They adapted
these methods to the unsupervised setting by first obtaining phone-like units
with the \gls{dpgmm}, an unsupervised clustering technique, and then using the
cluster assignments as unsupervised phone labels during \gls{asr} training.

In this work we instead consider a very simple feature-space adaptation method,
\acrfull{vtln}, which normalizes a speaker's speech by warping the
frequency-axis of the spectra. \gls{vtln} models are trained using maximum
likelihood estimation under a given acoustic model---here, an unsupervised
\gls{gmm}. Warp factors can then be extracted for both the training data and for
unseen data.

Although VTLN has recently been used by a few zero-resource speech
systems~\citep{Chen2015,Heck2017,Heck2018}, its impact in these systems is
unclear because there is no comparison to a baseline without \gls{vtln}.
\citet{Chen2017} did precisely such a comparison and showed that applying
\gls{vtln} to the input of their unsupervised feature learning method improved
its results in a phoneme discrimination task, especially in the cross-speaker
case. However, we don't know whether other feature learning methods are
similarly benefited by \gls{vtln}, nor even how \gls{vtln} on its own performs
in comparison to more recent methods. Thus, our first set of experiments is
designed to answer these questions by evaluating the benefits of using
\gls{vtln} and \gls{cae} learning, both on their own and in combination.

\subsection{Experimental Setup}
\label{sec:unsup_setup}

We use the GlobalPhone corpus of speech read from news
articles~\citep{Schultz2013}. We chose 6~languages from different language
families as \textbf{zero-resource} languages on which we evaluate the new
feature representations. That means our models do not have any access to the
transcriptions of the training data, although transcriptions still need to be
available to run the evaluation. The selected languages and dataset sizes are
shown in Table~\ref{tab:data-zero}. Each GlobalPhone language has recordings
from around 100~speakers, with 80\% of these in the training sets and no speaker
overlap between training, development, and test sets.

\begin{table}[th]
  \caption{Zero-resource languages, dataset sizes in hours.}
  \label{tab:data-zero}
  \centering
  \begin{tabular}{l l c c c}
    \toprule
    \multicolumn{2}{c}{\textbf{Language}} & 
    \multicolumn{1}{c}{\textbf{Train}} &
    \multicolumn{1}{c}{\textbf{Dev}} &
    \multicolumn{1}{c}{\textbf{Test}} \\
    \midrule
    \textit{GlobalPhone}\\
    Croatian   & (HR) & 12.1 & 2.0 & 1.8 \\
    Hausa      & (HA) &  6.6 & 1.0 & 1.1 \\
    Mandarin   & (ZH) & 26.6 & 2.0 & 2.4 \\
    Spanish    & (ES) & 17.6 & 2.1 & 1.7 \\
    Swedish    & (SV) & 17.4 & 2.1 & 2.2 \\
    Turkish    & (TR) & 13.3 & 2.0 & 1.9 \\
    \midrule
    \textit{ZRSC 2015} \\
    Buckeye English & (EN-B) & & & 5 \\
    Xitsonga        & (TS) & & & 2.5 \\
    \bottomrule
  \end{tabular}
\end{table}

For baseline features, we use Kaldi~\citep{Povey2011} to extract
MFCCs+$\Delta$+$\Delta\Delta$ and \gls{plps}+$\Delta$+$\Delta\Delta$ with a window
size of 25~ms and a shift of 10~ms, and we apply per-speaker \acrlong{cmn}. We
also evaluated MFCCs and PLPs with \gls{vtln}. The acoustic model used to
extract the warp factors was a diagonal-covariance \gls{gmm} with
1024~components. A single GMM was trained unsupervised on each language's
training data.

To train the \gls{cae}, we obtained \gls{utd} pairs using a freely available \gls{utd}
system\footnote{https://github.com/arenjansen/ZRTools}~\citep{Jansen2011} and
extracted 36k word pairs for each target language. Published results with this
system use PLP features as input, and indeed our preliminary experiments
confirmed that MFCCs did not work as well. We therefore report results using
only PLP or PLP+VTLN features as input to \gls{utd}.
Following~\citet{Renshaw2015} and \citet{Kamper2015}, we train the \gls{cae}
model\footnote{https://github.com/kamperh/speech\_correspondence} by first
pre-training an autoencoder with eight 100-dimensional layers and a final layer
of size 39 layer-wise on the entire training data for 5~epochs with a learning
rate of $2.5\times10^{-4}$. We then fine-tune the network with same-word pairs
as weak supervision for 60~epochs with a learning rate of $2.5\times10^{-5}$.
Frame pairs are presented to the \gls{cae} using either MFCC, MFCC+VTLN, or BNF
representation, depending on the experiment (preliminary experiments indicated
that PLPs performed worse than MFCCs, so MFCCs are used as the stronger
baseline). Features are extracted from the final hidden layer of the \gls{cae}
as shown in Figure~\ref{f:cae}.

To provide an upper bound on \gls{cae} performance, we also report results using
{\em gold standard} same-word pairs for \gls{cae} training. As in
\citet{Kamper2015,Jansen2013,Yuan2017a}, we force-align the target language data
and extract all the same-word pairs that are at least 5~characters and
0.5~seconds long (between 89k and 102k pairs for each language).

\subsection{Evaluation}
\label{sec:evaluation}
All experiments in this section are evaluated using the same-different
task~\citep{Carlin2011}, which tests whether a given speech representation can
correctly classify two speech segments as having the same word type or not. For
each word pair in a pre-defined set $S$ the \gls{dtw} cost between the acoustic
feature vectors under a given representation is computed. Two segments are then
considered a match if the cost is below a threshold. Precision and recall at a
given threshold $\tau$ are defined as
$$ P(\tau) = \frac{M_{\mathrm{SW}}(\tau)}{M_{\text{all}}(\tau)},~~~
R(\tau) = \frac{M_{\mathrm{SWDP}}(\tau)}{|S_{\mathrm{SWDP}}|}$$ where $M$ is the
number of \gls{sw}, \gls{swdp} or all discovered matches at that threshold and
$|S_{\mathrm{SWDP}}|$ is the number of actual \gls{swdp} pairs in $S$. We can
compute a precision-recall curve by varying $\tau$. The final evaluation metric
is the \gls{ap} or the area under that curve. We generate evaluation sets of
word pairs for the GlobalPhone development and test sets from all words that are
at least 5~characters and 0.5~seconds long, except that we now also include
different-word pairs.

Previous work~\citep{Carlin2011,Kamper2015} calculated recall with all \gls{sw}
pairs for easier computation because their test sets included a negligible
number of \gls{swsp} pairs. In our case the smaller number of speakers in the
GlobalPhone corpora results in up to 60\% of \gls{sw} pairs being from the same
speaker. We therefore always explicitly compute the recall only for \gls{swdp}
pairs to focus the evaluation of features on their speaker invariance.

\subsection{Results and Discussion}

Table~\ref{tab:vtln-results} shows AP results on all target languages for
\gls{cae} features learned using raw features as input (as in previous work) and
for \gls{cae} features learned using \gls{vtln}-adapted features as input to
either the \gls{utd} system, the \gls{cae}, or both. Baselines are raw MFCCs, or
MFCCs with VTLN. MFCCs with VTLN have not previously been compared to more
recent unsupervised subword modeling methods, but as our results show, they are
a much stronger baseline than MFCCs alone. Indeed, they are nearly as good as
\gls{cae} features (as trained in previous work). However, we obtain much better
results by applying \gls{vtln} to both the \gls{cae} and \gls{utd} input
features (MFCCs and PLPs, respectively). Individually these changes each result
in substantial improvements that are consistent across all 6 languages, and
applying VTLN at both stages helps further. Indeed, applying \gls{vtln} is
beneficial even when using gold pairs as \gls{cae} input, although to a lesser
degree.

So, although previous studies have indicated that cAE training and VTLN are
helpful individually, our experiments provide further evidence and
quantification of those results. In addition, we have shown that combining the
two methods leads to further improvements, suggesting that \gls{cae} training
and \gls{vtln} abstract over different aspects of the speech signal and should
be used together. The large gains we found with VTLN, and the fact that it was
part of the winning system in the 2017 ZRSC, suggest that it is also likely to
help in combination with other unsupervised subword modeling methods.

\begin{table}[t]
    \caption{Average precision scores on the same-different task (dev sets),
      showing the effects of applying \gls{vtln} to the input features for the
      \gls{utd} and/or \gls{cae} systems. \gls{cae} input is either MFCC or
      MFCC+VTLN. Topline results (rows 5-6) train cAE on gold standard pairs,
      rather than UTD output. Baseline results (final rows) directly evaluate
      acoustic features without UTD/cAE training. Best unsupervised result in
      bold.}
\label{tab:vtln-results}
  \centering
  \begin{tabular}{@{~~}l l c@{~~~~}c@{~~~~}c@{~~~~}c@{~~~~}c@{~~~~}c@{~~}}
    \toprule
    \multicolumn{1}{c}{\parbox{1cm}{\textbf{UTD\\ input}}} &
    \multicolumn{1}{c}{\parbox{.7cm}{\textbf{cAE \\input}}} & 
    \textbf{ES} &
    \textbf{HA} &
    \textbf{HR} &
    \textbf{SV} &
    \textbf{TR} &
    \textbf{ZH}\\
    \midrule
    \multicolumn{2}{l}{\quad\quad\textit{cAE systems:}} & \\
    PLP & MFCC      & 28.6 & 39.9 & 26.9 & 22.2 & 25.2 & 20.4 \\
    PLP & MFCC+VTLN & 46.2 & 48.2 & 36.3 & 37.9 & 31.4 & 35.7 \\
    \addlinespace[0.5em]
    PLP+VTLN & MFCC      & 40.4 & 45.7 & 35.8 & 25.8 & 25.9 & 26.9 \\
    PLP+VTLN & MFCC+VTLN & \textbf{51.5} & \textbf{52.9} & \textbf{39.6} & \textbf{42.9} & \textbf{33.4} & \textbf{44.4} \\
    \midrule
    \textit{Gold pairs} & MFCC      & 65.3 & 65.2 & 55.6 & 52.9 & 50.6 & 60.5 \\
    \textit{Gold pairs} & MFCC+VTLN & 68.9 & 70.1 & 57.8 & 56.9 & 56.3 & 69.5 \\
    \midrule
    \midrule
    \multicolumn{2}{@{~~}l}{\textit{Baseline:} MFCC}
    & 18.3 & 19.6 & 17.6 & 12.3 & 16.8 & 18.3 \\
    \multicolumn{2}{@{~~}l}{\textit{Baseline:} MFCC+VTLN}
    & 27.4 & 28.4 & 23.2 & 20.4 & 21.3 & 27.7 \\
    \bottomrule
  \end{tabular}
\end{table}

\section{Supervision from High-Resource Languages}\label{s:bnfs}

Next we investigate how labeled data from high-resource languages can be used
to obtain improved features on a target zero-resource language for which
no labeled data is available. Furthermore, are there benefits in deriving
this weak supervision from multiple languages?

\subsection{Background and Motivation}
There is considerable evidence that \gls{bnfs} extracted using a multilingually
trained \gls{dnn} can improve ASR for target languages with just a few hours of
transcribed data \citep{Vesely2012,Vu2012,Thomas2012,Cui2015,Alumae2016}.
However, there has been little work so far exploring supervised multilingual
\gls{bnfs} for target languages with no transcribed data at all.
\citet{Yuan2016} and \citet{Renshaw2015} trained {\em monolingual} BNF
extractors and showed that applying them cross-lingually improves word
discrimination in a zero-resource setting. \citet{Yuan2017} and \citet{Chen2017}
trained a multilingual \gls{dnn} to extract BNFs for a zero-resource task, but
the \gls{dnn} itself was trained on untranscribed speech: an unsupervised
clustering method was applied to each language to obtain phone-like units, and
the \gls{dnn} was trained on these unsupervised phone labels.

We know of only two previous studies of supervised multilingual BNFs for
zero-resource speech tasks. In the first, \citet{Yuan2017a} trained \gls{bnfs}
on either Mandarin, Spanish or both, and used the trained \gls{dnn}s to extract
features from English (simulating a zero-resource language). On a
query-by-example task, they showed that \gls{bnfs} always performed better than
MFCCs, and that bilingual \gls{bnfs} performed as well or better than
monolingual ones. Further improvements were achieved by applying weak
supervision in the target language using a \gls{cae}
trained on English word pairs. However, the authors did not
experiment with more than two training languages, and only evaluated on English.

In the second study, \citet{Shibata2017} built multilingual systems
using either seven or ten high-resource languages, and evaluated on the three
``development'' and two ``surprise'' languages of the \gls{zrsc} 2017. However,
they included transcribed training data from four out of the five evaluation
languages, so only one language's results (Wolof) were
truly zero-resource.

Our experiments therefore aim to evaluate on a wider range of target languages,
and to explore the effects of both the {\em amount} of labeled data, and the
{\em number of languages} from which it is obtained.

\subsection{Experimental Setup}
We picked another 10~languages (different from the target languages described in
Section~\ref{sec:unsup_setup}) with a combined 198.3~hours of speech from the
GlobalPhone corpus. We consider these as \textbf{high-resource} languages, for
which transcriptions are available to train a supervised \gls{asr} system. The
languages and dataset sizes are listed in Table~\ref{tab:data-high}. We also use
the English \gls{wsj} corpus~\citep{Paul1992} which is comparable to the
GlobalPhone corpus. It contains a total of 81~hours of speech, which we either
use in its entirety or from which we use a 15~hour subset; this allows us to
compare the effect of increasing the amount of data for one language with
training on similar amounts of data but from different languages.

\begin{table}
  \caption{High-resource languages, dataset sizes in hours.}
  \label{tab:data-high}
  \centering
  \begin{tabular}{l l c}
    \toprule
    \multicolumn{2}{c}{\textbf{Language}} & 
    \multicolumn{1}{c}{\textbf{Train}}\\
    \midrule
    Bulgarian  & (BG) & 17.1 \\
    Czech      & (CS) & 26.8 \\
    French     & (FR) & 22.8 \\
    German     & (DE) & 14.9 \\
    Korean     & (KO) & 16.6 \\
    Polish     & (PL) & 19.4 \\
    Portuguese & (PT) & 22.8 \\
    Russian    & (RU) & 19.8 \\
    Thai       & (TH) & 21.2 \\
    Vietnamese & (VI) & 16.9 \\
    \midrule
    English81 WSJ & (EN) & 81.3 \\
    English15 WSJ &      & 15.1 \\
    \bottomrule
  \end{tabular}
\end{table}

Supervised models trained on these high-resource languages are evaluated on the
same set of zero-resource languages as in Section~\ref{sec:unsupervised}.
Transcriptions of the latter are still never used during training.

For initial monolingual training of \gls{asr} systems for the high-resource
languages, we follow the Kaldi recipes for the GlobalPhone and WSJ corpora and
train a \gls{sgmm} system for each language to get initial context-dependent
state alignments; these states serve as targets for \gls{dnn} training.

\begin{figure}[t!]
\centering
  \includestandalone[width=0.5\columnwidth]{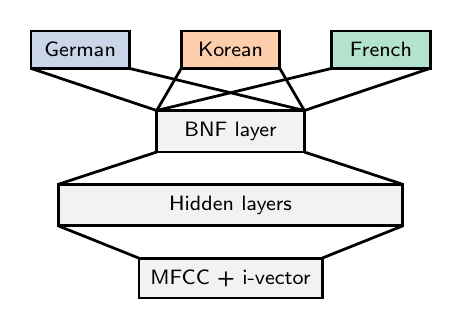}
  \caption{Multilingual acoustic model training architecture. All layers are
    shared between languages except for the language-specific output layers at
    the top.}
  \label{fig:bnf-training}
\end{figure}

For multilingual training, we closely follow the existing Kaldi recipe for the
Babel corpus~\citep{Trmal2017a}. We train a
\gls{tdnn}~\citep{Waibel1989,Lang1990,Peddinti2015} with block
softmax~\citep{Grezl2014}, i.e. all hidden layers are shared between languages,
but there is a separate output layer for each language. For each training
instance only the error at the corresponding language's output layer is used to
update the weights. This architecture is illustrated in
Figure~\ref{fig:bnf-training}. The \gls{tdnn} has six 625-dimensional hidden
layers\footnote{The splicing indexes are \texttt{-1,0,1 -1,0,1 -1,0,1 -3,0,3
    -3,0,3 -6,-3,0 0}.} followed by a 39-dimensional bottleneck layer with ReLU
activations and batch normalization. Each language then has its own
625-dimensional affine and a softmax layer. The inputs to the network are
40-dimensional MFCCs with all cepstral coefficients to which we append i-vectors
for speaker adaptation. The network is trained with stochastic gradient descent
for 2~epochs with an initial learning rate of $10^{-3}$ and a final learning
rate of $10^{-4}$.

In preliminary experiments we trained a separate i-vector extractor for each
different sized subset of training languages. However, results were similar to
training on the pooled set of all 10 high-resource languages, so for expedience
we used the 100-dimensional i-vectors from this pooled training for all reported
experiments. The i-vectors for the zero-resource languages are obtained from the
same extractor. This allows us to also apply speaker adaptation in the
zero-resource scenario and to draw a fair comparison with our best cAE
  results that made use of speaker normalisation in the form of \gls{vtln}.
  The results will only show the effect of increasing the number of training
languages because the acoustic models are always trained with i-vectors.
Including i-vectors yielded a small performance gain over not doing so; we also
tried applying \gls{vtln} to the MFCCs for \gls{tdnn} training, but found no
additional benefit.

\subsection{Results and Discussion}

\begin{table}[b]
  \caption{Word error rates of monolingual \gls{sgmm} and 10-lingual TDNN ASR system
    evaluated on the development sets.}
  \label{tab:wer-results}
  \centering
  \begin{tabular}{l c c c l c c}
    \toprule
    \multicolumn{1}{c}{\textbf{Language}} & 
    \multicolumn{1}{c}{\textbf{Mono}} &
    \multicolumn{1}{c}{\textbf{Multi}} &\quad\quad\quad\quad&
    \multicolumn{1}{c}{\textbf{Language}} & 
    \multicolumn{1}{c}{\textbf{Mono}} &
    \multicolumn{1}{c}{\textbf{Multi}} \\
    \midrule
    BG & 17.5 & 16.9 && PL & 16.5 & 15.1 \\
    CS & 17.1 & 15.7 && PT & 20.5 & 19.9 \\
    DE & 9.6 & 9.3 && RU & 27.5 & 26.9 \\
    FR & 24.5 & 24.0 && TH & 34.3 & 33.3 \\
    KO & 20.3 & 19.3 && VI & 11.3 & 11.6 \\
    \bottomrule
  \end{tabular}
\end{table}

\begin{figure*}[t!]
  \includestandalone[width=\textwidth]{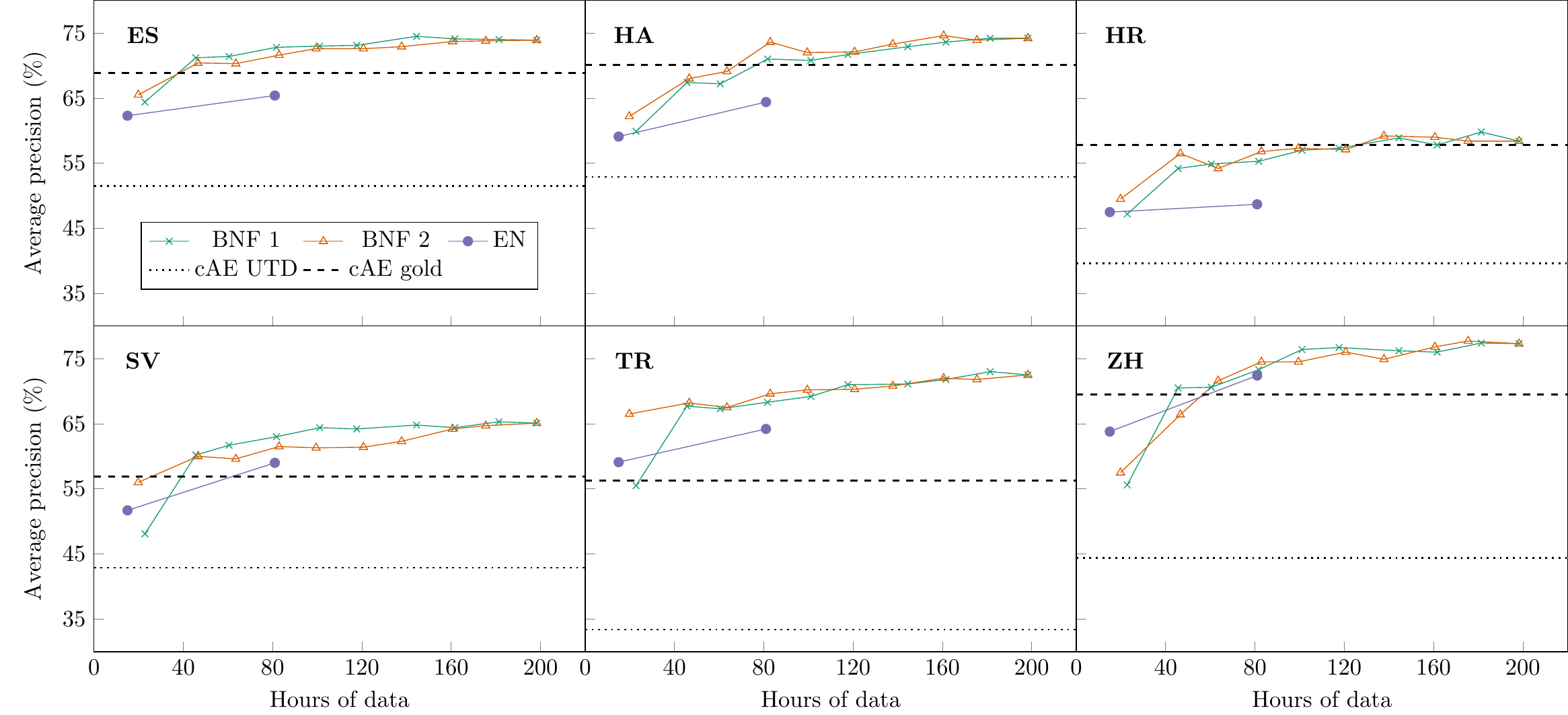}
  \caption{Same-different task evaluation on the development sets for BNFs
    trained on different amounts of data. We compare training on up to
    10~different languages with additional data in one language (English). For
    multilingual training, languages were added in two different orders:
    FR-PT-DE-TH-PL-KO-CS-BG-RU-VI (BNFs 1) and RU-CZ-VI-PL-KO-TH-BG-PT-DE-FR
    (BNFs 2). Each datapoint shows the result of adding an additional language.
    As baselines we include the best unsupervised cAE and the cAE trained on
    gold standard pairs from rows 4 and 6 of Table~\ref{tab:vtln-results}.}
  \label{fig:bnf-results}
\end{figure*}

As a sanity check we include word error rates (WER) for the acoustic models
trained on the high-resource languages. Table~\ref{tab:wer-results} compares the
WER of the monolingual \gls{sgmm} systems that provide the targets for
\gls{tdnn} training to the WER of the final model trained on all
10~high-resource languages. The multilingual model shows small but consistent
improvements for all languages except Vietnamese. Ultimately though, we are not
so much interested in the performance on typical \gls{asr} tasks, but in whether
\gls{bnfs} from this model also generalize to zero-resource applications on
unseen languages.

Figure~\ref{fig:bnf-results} shows \gls{ap} on the same-different task of
multilingual \gls{bnfs} trained from scratch on an increasing number of
languages in two randomly chosen orders.
We provide two baselines for
comparison, drawn from our results in Table~\ref{tab:vtln-results}. Firstly, our
best \gls{cae} features trained with \gls{utd} pairs (row 4,
Table~\ref{tab:vtln-results}) are a reference for a fully unsupervised system.
Secondly, the best \gls{cae} features trained with gold standard pairs (row
6, Table~\ref{tab:vtln-results}) give an upper bound on the \gls{cae}
performance.

In all 6~languages, even \gls{bnfs} from a monolingual \gls{tdnn} already
considerably outperform the \gls{cae} trained with \gls{utd} pairs. Adding
another language usually leads to an increase in \gls{ap}, with the \gls{bnfs}
trained on 8--10~high-resource languages performing the best, also always
beating the gold \gls{cae}. The biggest performance gain is obtained from adding
a second training language---further increases are mostly smaller. The order of
languages has only a small effect, although for example adding other Slavic
languages is generally associated with an increase in \gls{ap} on Croatian. This
suggests that it may be beneficial to train on languages related to the
zero-resource language if possible, but further experiments need to be conducted
to quantify this effect.

To determine whether these gains come from the diversity of training languages
or just the larger amount of training data, we trained models on the 15~hour
subset and the full 81~hours of the English \gls{wsj} corpus, which corresponds
to the amount of data of four GlobalPhone languages. More data does help to some
degree, as Figure~\ref{fig:bnf-results} shows. But, except for Mandarin, training
on just two languages (46~hours) already works better.

\section{Evaluation using ZRSC Data and Measures}

Do the results from the previous experiments generalise to other corpora and
how do they compare to other works? So far we used data from GlobalPhone, which
provides corpora collected and formatted similarly for a wide range of
languages. However, GlobalPhone is not freely available and no previous
zero-resource studies have used these corpora, so in this section we also
provide results on the \acrfull{zrsc} 2015~\citep{Versteegh2015} data sets,
which have been widely used in other work. The target languages that we
treat as zero-resource are English (from the Buckeye corpus~\citep{buckeye07})
and Xitsonga (NCHLT corpus~\citep{devries+etal_speechcom14}).
Table~\ref{tab:data-zero} includes the statistics of the subsets of these
corpora that were used in the \gls{zrsc} 2015 and in this work. These corpora
are not split into train/dev/test; since training is unsupervised, the system is
simply trained directly on the unlabeled test set (which could also be done in
deployment). Importantly, no hyperparameter tuning is done on the Buckeye or
Xitsonga data, so these results still provide a useful test of generalization.
Notably, the Buckeye English corpus contains conversational speech and is
therefore different in style from the rest of our data.

For training the \gls{cae} on the Buckeye English and
Xitsonga corpora, we use the same sets of \gls{utd} pairs as
\citet{Renshaw2015}, which were discovered from \gls{fdlp} features.
We evaluate using both the same-different measures from above, as well as 
the ABX phone discriminability task~\citep{Schatz2013}
used in the \gls{zrsc} and other recent work~\citep{Versteegh2015,Dunbar2017}.
The ABX task evaluates phoneme discriminability using minimal pairs: sequences
of three phonemes where the central phoneme differs between the two sequences
$A$ and $B$ in the pair, such as \texttt{b ih n} and \texttt{b eh n}. 
  Feature representations are then evaluated on how well they can identify a
  third triplet $X$ as having the same phoneme sequence as either $A$ or $B$.
  See \citet{Versteegh2015} and \citet{Dunbar2017} for details on how the scores are computed
  and averaged over speakers and phonemes to obtain the final ABX error rate.
  One usually distinguishes between the \textit{within-speaker} error rate where
  all three triplets belong to the same speaker, and the \textit{cross-speaker}
  error rate where $A$ and $B$ are from the same and $X$ from a different
  speaker. 

The ABX evaluation includes all such minimal pair phoneme triplets of the
  evaluation corpus. These pairs therefore rarely correspond to full words,
  making it a somewhat abstract task whose results may be difficult to interpret
  when summarizing it as a single final metric. ABX can however be very suitable
  for more fine-grained analysis of speech phenomena by including only specific
  phonetic contrasts in the evaluation~\citep{Schatz2018}. In contrast, the same-different
  task always compares whole words and directly evaluates how
  good feature representations are at telling whether two utterances are the
  same word or not. Thus it has an immediate link to applications like spoken
  term detection and it allows easier error analysis. It is also faster to
  prepare the same-different evaluation set and run the evaluation. We wish
  to verify that the ABX and same-different measures correlate well, to better compare
  studies that use only one of them and to allow choosing the task that is more
  appropriate for the situation at hand.

\begin{table}[t]
  \caption{Comparison of AP on the same-different task (higher is better) and
    ABX cross-/within-speaker error rates (lower is better) for the Buckeye English and Xitsonga
    corpora.}
  \label{tab:bucktsonga}
  \centering
  \begin{tabular}{l c c@{~~}c@{~~~~}c@{~~}c}
    \toprule
    && \multicolumn{2}{c}{English} & \multicolumn{2}{c}{Xitsonga} \\
    \cmidrule(lr){3-4}
    \cmidrule(l){5-6}
    \textbf{Features} & Dimensions & ABX & Same-diff & ABX & Same-diff \\
    \midrule
    \multicolumn{5}{l}{\textit{Unsupervised}}\\
    MFCC & 39 & 28.4 / 15.5 & 19.14 & 33.4 / 20.9 & 10.46 \\
    MFCC+VTLN & 39 & 26.5 / 15.4 & 24.19 & 31.9 / 21.4 & 13.33 \\
    cAE & 39 & 24.0 / 14.5 & 31.97 & 23.8 / 14.8 & 22.79 \\
    cAE+VTLN & 39 & 22.9 / 14.3 & 37.85 & 22.6 / 14.5 & 47.41 \\
    \\
    \multicolumn{5}{l}{\textit{Weak multilingual supervision}}\\
    BNF & 39 & 18.0 / 12.4 & 60.19 & 17.0 / 12.3 & 63.44 \\
    \midrule
    ZRSC Topline \citep{Versteegh2015} & 49 & 16.0 / 12.1 & - & 4.5 / 3.5 & - \\
    \citet{Heck2018} & 139--156 & 14.9 / 10.0 & - & 11.7 / 8.1 & - \\
    \citet{Riad2018} & 100 & 17.2 / 10.4 & - & 15.2 / 9.4 & - \\
    \bottomrule
  \end{tabular}
\end{table}

Table~\ref{tab:bucktsonga} shows results on the Xitsonga and Buckeye English
corpora. Here we compare ABX error rates computed with the \gls{zrsc}
2015~\citep{Versteegh2015} evaluation scripts with \gls{ap} on the same-different
task. To the best of our knowledge, this is the first time such a comparison has
been made. The results on both tasks correlate well, especially when looking at
the ABX cross-speaker error rate because the same-different evaluation as
described in Section~\ref{sec:evaluation} also focuses on cross-speaker pairs.
As might be expected \gls{vtln} only improves cross-speaker, but not
within-speaker ABX error rates.

For comparison we also include ABX results of the official \gls{zrsc} 2015
topline~\citep{Versteegh2015}, which are posteriorgrams obtained from a
supervised speech recognition system, the current state-of-the-art
system\footnote{The ZRSC website maintains a list of results: https://zerospeech.com/2015/results.html}~\citep{Heck2018} which even outperforms the topline for English, and
the system of~\citet{Riad2018} which is the most recent form of the 
ABNet~\citep{Synnaeve2014}, an architecture that is similar to our \gls{cae}.

These systems score better than all of our features, but are not directly
comparable for several reasons. Firstly, it is unclear how these systems were
optimized, since there was no separate development set in \gls{zrsc} 2015.
Secondly, our features are all 39-dimensional to be directly comparable with
MFCCs, whereas the other two systems have higher dimensionality---indeed, the
dimensionality of the winning \gls{dpgmm} system from the \gls{zrsc} 2017 was even
greater, with more than 1000~dimensions~\citep{Heck2017}---and we don't know
whether the competing systems would work as well with fewer dimensions. Such
higher dimensional features may be useful in some circumstances, but require
more memory and processing power to use, which could be undesirable or even
prohibitive for some downstream applications (such as the unsupervised
segmentation and clustering system used in Section~\ref{sec:segmentation}.
\citet{Heck2018a} propose a more complex sampling approach to reduce the
potentially very high dimensionality of the features obtained with the
\glspl{dpgmm} in their previous work. However, the resulting dimensionality is
still around 90--120 and cannot be controlled precisely, which might be required
for down-stream applications.

  This complexity of evaluating zero-resource subword modeling systems was
  addressed in the \gls{zrsc} 2019, where the bitrate of the features was added as
  another evaluation metric alongside the ABX performance~\citep{Dunbar2019}. This
  means that systems are now compared on two dimensions and researchers may choose
  to trade off between the two, while ultimately the goal is to find a representation
  that performs well on both measures, like phonemes.

The \gls{bnfs} are in any case competitive with the higher dimensional features,
and have the advantage that they can be built using standard Kaldi scripts and
do not require any training on the target language, so can easily be deployed to
new languages. The competitive result of~\citet{Riad2018} also shows that in
general a system trained on word pairs discovered from a \gls{utd} system can
perform very well.

\section{Can We Improve the Multilingual BNFs?}

So far we have shown that multilingual \gls{bnfs} that are completely agnostic
to the target language work better than any of the features trained using only
the target language data. However, in principle it could be possible to improve
performance further by passing the \gls{bnfs} as inputs to models that train on
the target language data in an unsupervised fashion. We explored this
possibility by simply training a \gls{cae} using \gls{bnfs} as input rather than
MFCCs. That is, we trained the \gls{cae} with the same word pairs as before, but
replaced VTLN-adapted MFCCs with the 10-lingual \gls{bnfs} as input features,
without any other changes in the training procedure. Table~\ref{tab:cae-results}
(penultimate row) shows that the \gls{cae} trained with \gls{utd} pairs is able
to slightly improve on the \gls{bnfs} in some cases, but this is not consistent
across all languages and for Croatian the \gls{cae} features are much worse. On
the other hand, when trained using gold standard pairs (final row), the
resulting \gls{cae} features {\em are} consistently better than the input
\gls{bnfs}. This indicates that \gls{bnfs} can in principle be improved by
further unsupervised target-language training, but the top-down supervision
needs to be of higher quality than the current UTD system provides.

\begin{table}[b]
  \caption{\Gls{ap} on the same-different task when training
    \gls{cae} on the 10-lingual \gls{bnfs} from above (cAE-BNF) with UTD and
    gold standard word pairs (test set results). Baselines are MFCC+VTLN and the
    cAE models from rows 4 and 6 of Table~\ref{tab:vtln-results} that use
    MFCC+VTLN as input features. Best result without target language supervision
    in bold.}
  \label{tab:cae-results}
  \centering
  \begin{tabular}{l c@{~~~~}c@{~~~~}c@{~~~~}c@{~~~~}c@{~~~~}c@{}}
    \toprule
    \multicolumn{1}{c}{\textbf{Features}} & 
    \textbf{ES} &
    \textbf{HA} &
    \textbf{HR} &
    \textbf{SV} &
    \textbf{TR} &
    \textbf{ZH}\\
    \midrule
    MFCC+VTLN       & 44.1 & 22.3 & 25.0 & 34.3 & 17.9 & 33.4 \\
    cAE UTD         & 72.1 & 41.6 & 41.6 & 53.2 & 29.3 & 52.8 \\
    cAE gold        & 85.1 & 66.3 & 58.9 & 67.1 & 47.9 & 70.8 \\
    \midrule
    10-lingual BNFs & \textbf{85.3} & \textbf{71.0} & \textbf{56.8} & 72.0 & \textbf{65.3} & 77.5 \\
    \midrule
    cAE-BNF UTD     & 85.0 & 67.4 & 40.3 & \textbf{74.3} & 64.6 & \textbf{78.8} \\
    cAE-BNF gold    & 89.2 & 79.0 & 60.8 & 79.9 & 69.5 & 81.6 \\
    \bottomrule
  \end{tabular}
\end{table}

\begin{figure}[htb] \centering
\includestandalone[width=\columnwidth]{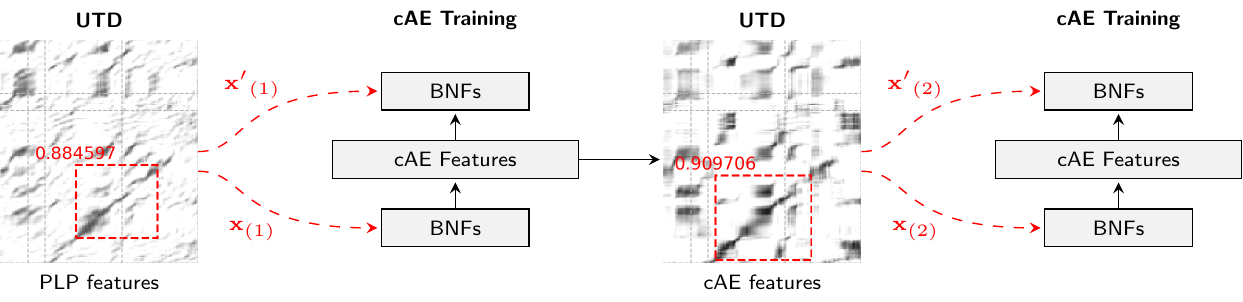}
\caption{Cycling cAE and UTD systems: Iteration 1) UTD is run on PLP
features to obtain word pairs. Pairs (\textbf{x}, \textbf{x'}) are then
represented using multilingual BNFs and used to train the cAE. Iteration 2)
Features from the last hidden cAE layer are passed to the UTD system, which
discovers new word pairs that can be used in the next iteration of cAE training.}
  \label{fig:cae-utd-cycle}
\end{figure}

This observation leads to a further question: could we improve the UTD pairs
themselves by using our improved features (either \gls{bnfs} or \gls{cae}
features) as input to the UTD system? If the output is a better set of UTD pairs
than the original set, these could potentially be used to further improve the
features, and perhaps the process could be iterated as illustrated in Figure
\ref{fig:cae-utd-cycle}. As far as we know, no previously published work has combined unsupervised
subword modeling with a \gls{utd} system.\footnote{While some other work, such
as \citet{Lee2015} and \citet{Walter2013}, has focused on joint phonological and
lexical discovery, these do not perform representation learning on the low-level
features.}

\begin{figure}[tb!]
  \centering
  \begin{subfigure}[b]{0.24\textwidth}
    \includegraphics[height=1.8in]{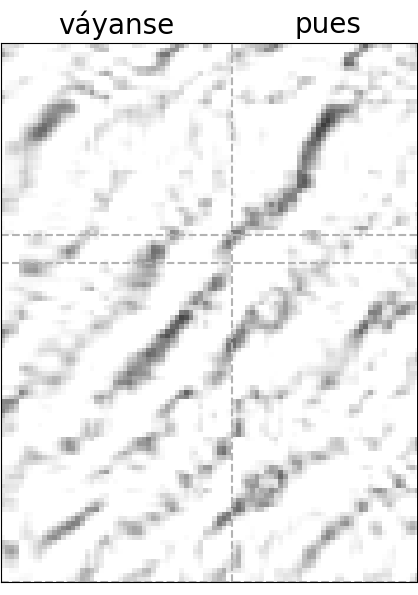}
    \caption{PLP}
  \end{subfigure}
  \begin{subfigure}[b]{0.24\textwidth}
    \includegraphics[height=1.8in]{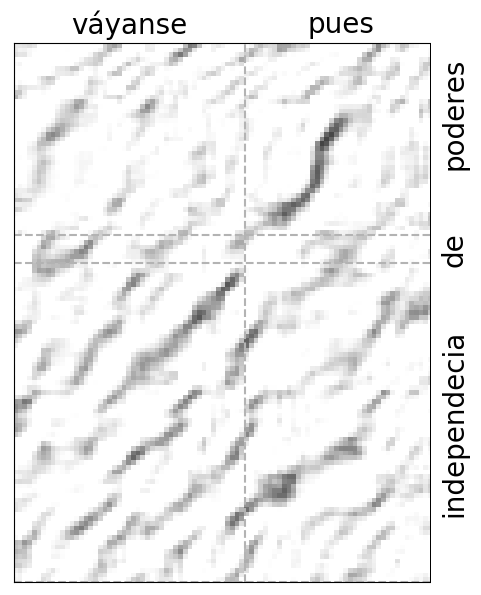}
    \caption{PLP-VTLN}
  \end{subfigure}
  \begin{subfigure}[b]{0.24\textwidth}
    \includegraphics[height=1.8in]{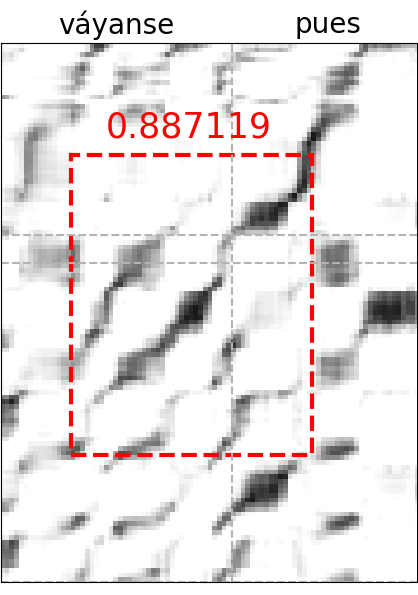}
    \caption{cAE UTD}
  \end{subfigure}
  \begin{subfigure}[b]{0.24\textwidth}
    \includegraphics[height=1.8in]{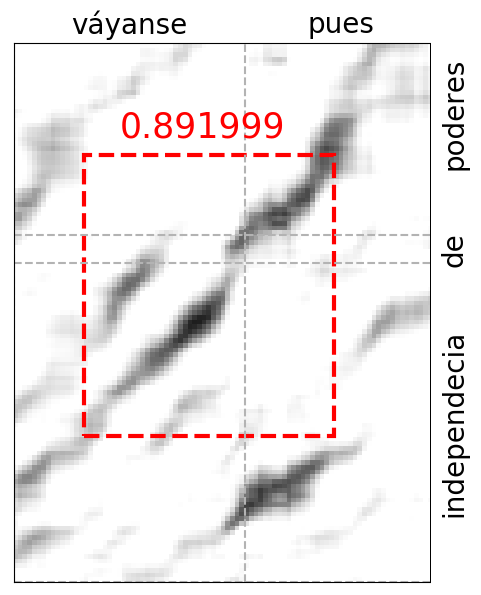}
    \caption{BNF}
  \end{subfigure}
  \caption{Frame-wise cosine similarity matrices for two Spanish utterances from
    different speakers, comparing different feature representations. Dark
    regions correspond to high cosine similarity and values below 0.4 are
    clipped. Red rectangles mark matches discovered by the UTD system and
    include their DTW similarity scores. The discovered matches are
    incorrect---although phonetically similar---and found only for cAE features
    and BNFs.}
  \label{fig:utd-alignments6}
\end{figure}

\begin{figure*}[!]
  \centering
  \begin{subfigure}[b]{0.47\textwidth}
    \includegraphics[width=3.3in]{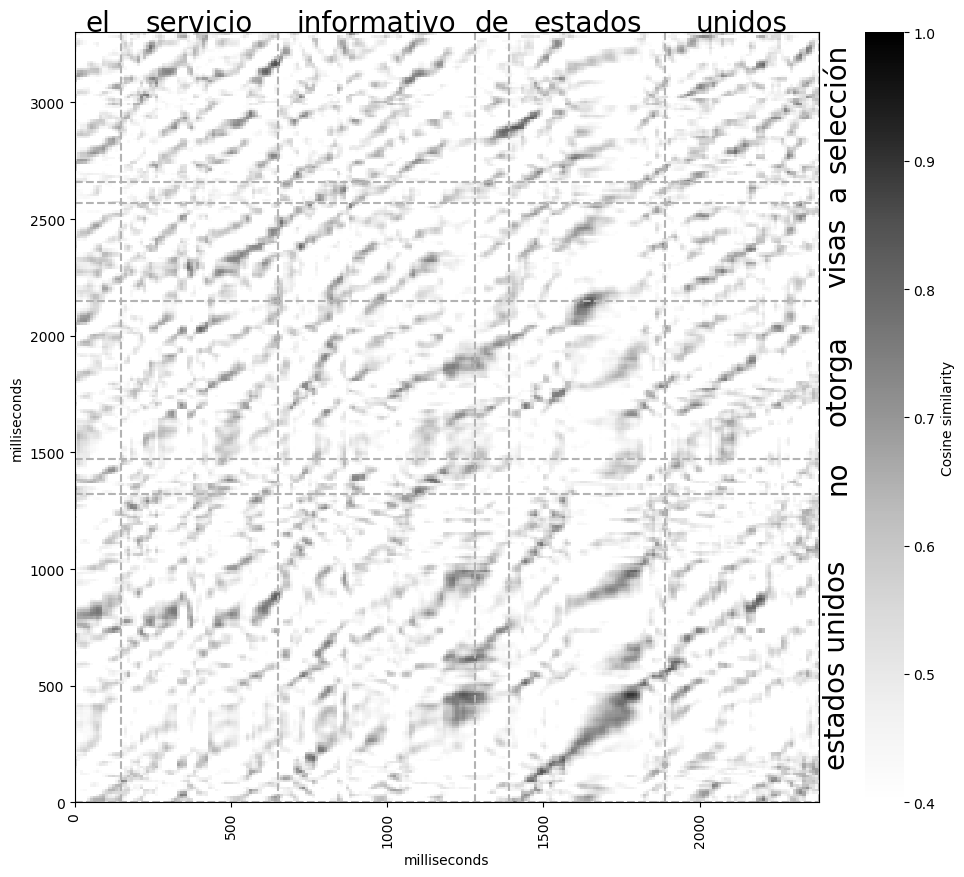}
    \caption{PLP}
  \end{subfigure}
  \quad
  \begin{subfigure}[b]{0.47\textwidth}
    \includegraphics[width=3.3in]{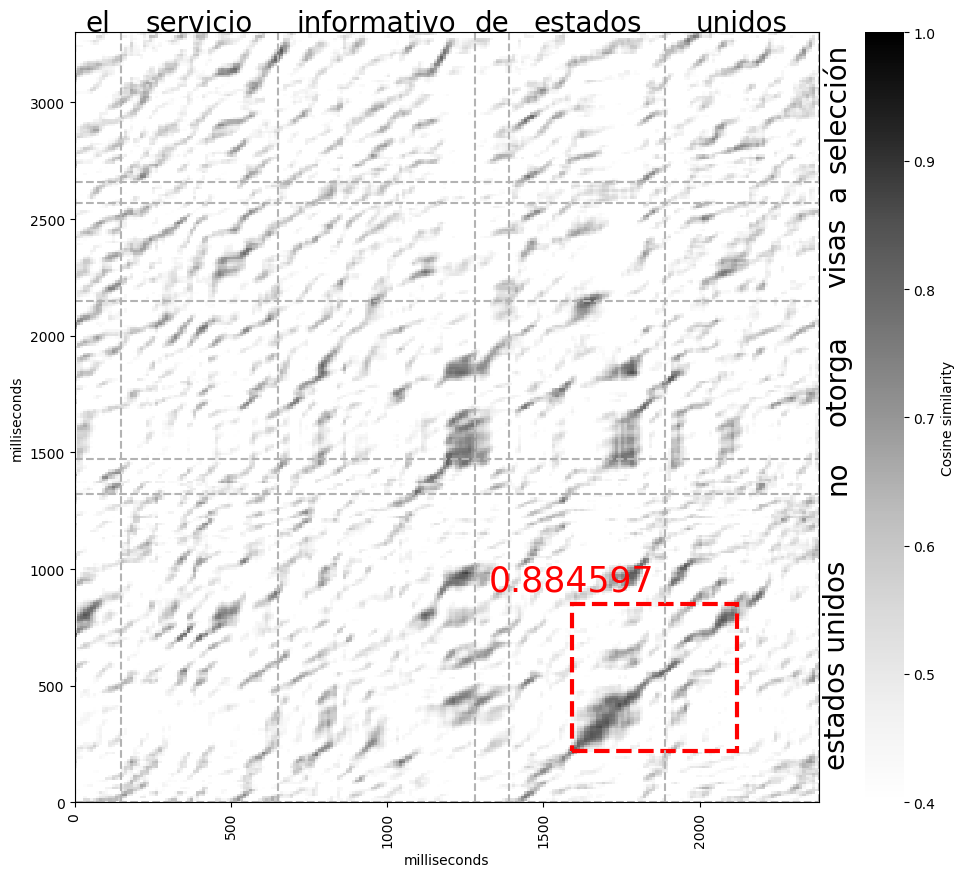}
    \caption{PLP-VTLN}
  \end{subfigure}
  \\
  \begin{subfigure}[b]{0.47\textwidth}
    \includegraphics[width=3.3in]{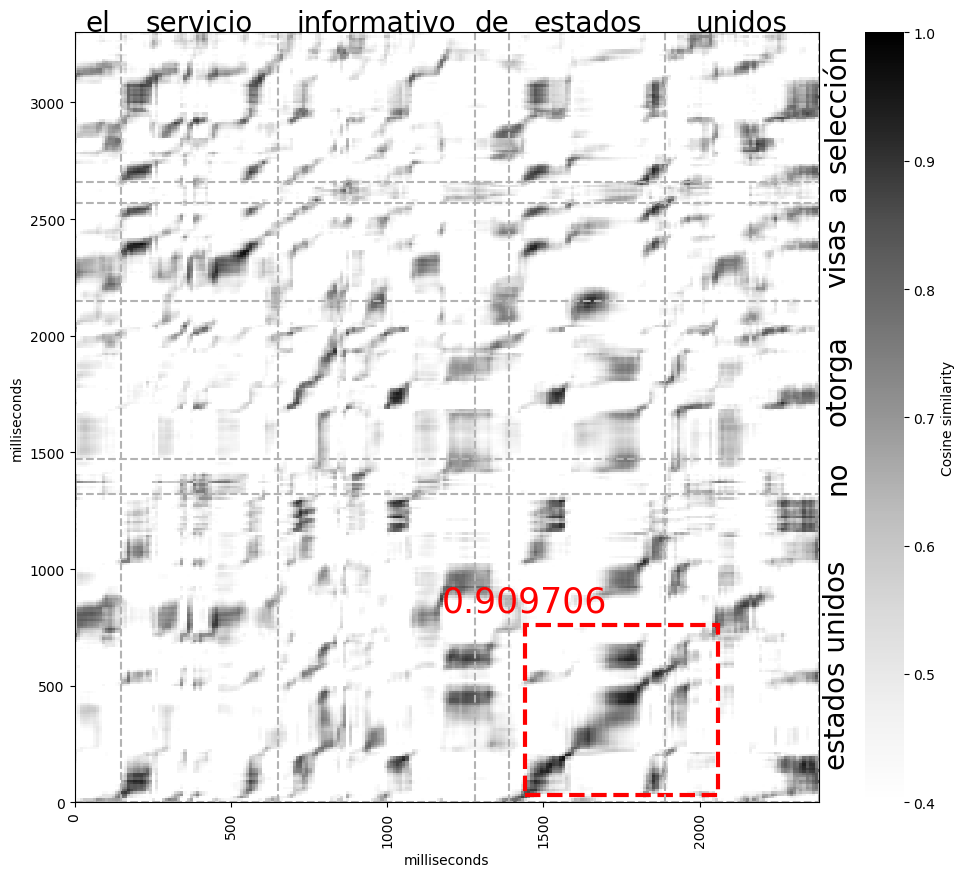}
    \caption{cAE UTD}
  \end{subfigure}
  \quad
  \begin{subfigure}[b]{0.47\textwidth}
    \includegraphics[width=3.3in]{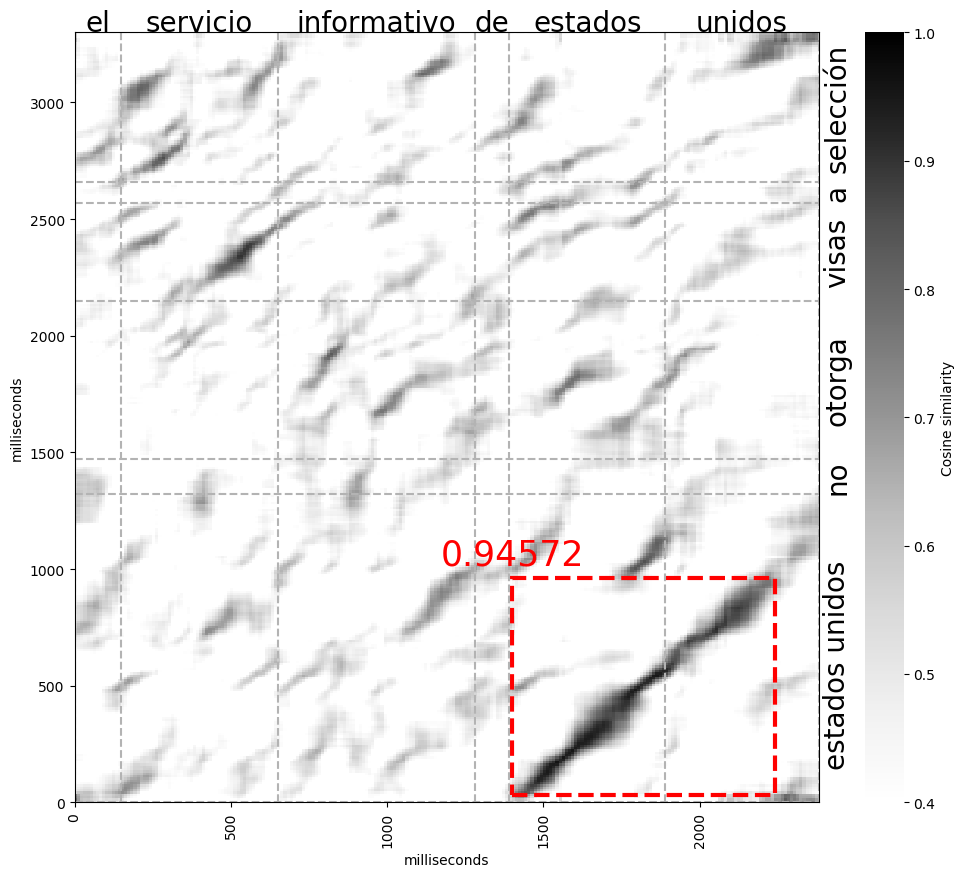}
    \caption{BNF}
  \end{subfigure}
  \\
  \caption{Frame-wise cosine similarity matrices for two Spanish utterances from
    different speakers, comparing different feature representations. Dark
    regions correspond to high cosine similarity and values below 0.4 are
    clipped. Red rectangles mark matches discovered by the UTD system
    and include their DTW similarity scores. In this case the match is not found
    with PLPs as input features.}
  \label{fig:utd-alignments5}
\end{figure*}

Unfortunately, after considerable effort to make this work we found that the
ZRTools \gls{utd} system seems to be too finely tuned towards features that
resemble PLPs to get good results from our new features. Examining the
similarity plots for several pairs of utterances helps explain the issue, and
also reveals interesting qualitative differences between our learned features
and the PLPs, as shown in Figures~\ref{fig:utd-alignments5} and
\ref{fig:utd-alignments6}. Darker areas in these plots indicate higher acoustic
similarity, so diagonal line segments point to similar sequences, as in
Figure~\ref{fig:utd-alignments5} where both utterances contain the words
\textit{estados unidos}. The ZRTools \gls{utd} toolkit identifies these diagonal
lines with fast computer vision techniques~\citep{Jansen2011} and then runs a
segmental-\gls{dtw} algorithm only in the candidate regions for efficient
discovery of matches.

PLPs are designed to contain fine-grained acoustic information about the speech
signal and can therefore vary a lot throughout the duration of a phoneme. Accordingly, the
diagonal lines in Figure~\ref{fig:utd-alignments5}~(a) are very thin
and there is a lot of spurious noise that does not necessarily correspond
to phonetically similar units. This pattern is similar for VTLN-adapted PLPs
in (b), but with less noise.

On the other hand, \gls{cae} features and \gls{bnfs} are  trained to
ignore such local variation within phonemes. This results in significantly
different appearance of frame-wise cosine similarity plots of two utterances.
The trained features remain more constant throughout the duration of a phoneme,
resulting in wider diagonal lines in the similarity plots. Especially \gls{cae}
features are very good at learning phoneme-level information, indicated by
the large rectangular blocks in Figure~\ref{fig:utd-alignments5}~(c) where
phonemes of the two utterances match or are very similar. We also found the
boundaries of these blocks to align well with actual phoneme boundaries
provided by forced alignment. This is despite the \gls{cae} not having any
information about phoneme identities or boundaries during training.

Parameters in the segmental \gls{dtw} algorithm of ZRTools, which searches
for exact matches within the identified diagonal line segments where matches are
likely to occur, include the maximum deviation of the path from the diagonal and
similarity budgets and thresholds that determine how far the path should extend
at each end~\citep{Jansen2011}. While ZRTools still finds the diagonal lines in \gls{cae}
features and \gls{bnfs}, the \gls{dtw} algorithm then finds too many
exact matches because the lines are much wider and similarity values overall
higher than for PLPs. For example Figure~\ref{fig:utd-alignments6} shows a
typical example of phonetically similar, but incorrect matches that are only
discovered in \gls{cae} features and \gls{bnfs}.

While this behaviour can be compensated for to some degree by changing
ZRTools' parameters, we could not identify metrics that can reliably predict
whether a given set of discovered UTD pairs will result in better cAE
performance. Thus tuning these parameters is very time-consuming because it
requires running both UTD and cAE training for each step. Although it might be
possible to eventually identify a set of \gls{dtw} parameters that can work with
features that are relatively stable within phones, it could be more productive
to consider different approaches for these types of features.

\section{Downstream Application: Segmentation and Clustering}
\label{sec:segmentation}
Our experiment with the UTD system was disappointing, suggesting that although
\gls{cae} features and \gls{bnfs} improve intrinsic discriminability measures,
they may not work with some downstream zero-resource tools. However, ZRTools is
a single example. To further investigate the downstream effects of the learned
features, we now consider the task of full-coverage speech segmentation and
clustering. The aim here is to tokenize the entire speech input into
hypothesized categories, potentially corresponding to words, and to do so
without any form of supervision---essentially a form of unsupervised speech
recognition. Such systems could prove useful from a speech technology
perspective in low-resource settings, and could be useful in studying how human
infants acquire language from unlabeled speech input.

Here we specifically investigate whether our BNFs improve the Bayesian
embedded segmental Gaussian mixture model (BES-GMM), first proposed by
\citet{Kamper2016}. This approach relies on a mapping where potential word
segments (of arbitrary length) are embedded in a fixed-dimensional acoustic
vector space. The model, implemented as a Gibbs sampler, builds a whole-word
acoustic model in this acoustic embedding space, while jointly performing
segmentation. Several acoustic word embedding methods have been considered, but
here we use the very simple approach also used \citet{Kamper2017}: any
segment is uniformly downsampled so that it is represented by the same fixed
number of frame-level features, which are then flattened to obtain the
fixed-dimensional embedding \citep{Levin2013}.

\subsection{Experimental Setup and Evaluation}
We retrained the \gls{cae} and BNF models to return 13-dimensional features with
all other parameters unchanged to be consistent with the experiments of
\citet{Kamper2017} and for computational reasons. We also did not tune any
hyperparameters of the BES-GMM for our new input features. Nonetheless, our
baseline \gls{cae} results do not exactly correspond to the ones of
\citet{Kamper2017} because for example the MFCC input features have been
extracted with a different toolkit and we used a slightly different training
procedure.

We use several metrics to compare the resulting segmented word tokens to ground
truth forced alignments of the data. By mapping every discovered word token to
the ground truth word with which it overlaps most, average \textbf{cluster
  purity} can be calculated as the total proportion of correctly mapped tokens
in all clusters. More than one cluster may be mapped to the same ground truth
word type. In a similar way, we can calculate \textbf{unsupervised word error
  rate (WER)}, which uses the same cluster-to-word mapping but also takes
insertions and deletions into account. Here we consider two ways to perform the
cluster mapping: many-to-one, where more than one cluster can be assigned the
same word label (as in purity), or one-to-one, where at most one cluster is
mapped to a ground truth word type (accomplished in a greedy fashion). We also
compute the \textbf{gender and speaker purity} of the clusters, where we want to
see clusters that are as diverse as possible on these measures, i.e., low
purity. To explicitly evaluate how accurate the model performs segmentation, we
compare the proposed word boundary positions to those from forced alignments of
the data (falling within a single true phoneme from the boundary). We calculate
boundary precision and recall, and report the resulting \textbf{word boundary
  F-scores}. We also calculate \textbf{word token F-score}, which requires that
both boundaries from a ground truth word token be correctly predicted.

\subsection{Results}
\begin{table*}[t]
  \caption{Segmentation and clustering results (lower scores are better,
    except for token and boundary F-score, and cluster purity).}
  \label{tab:segmentation}
  \centering
  \begin{tabular}{l c c c c c c c}
    \toprule
    & \multicolumn{2}{c}{\textbf{WER}} & \multicolumn{2}{c}{\textbf{F-score}} & \multicolumn{3}{c}{\textbf{Purity}} \\
    \cmidrule(lr){2-3} \cmidrule(lr){4-5} \cmidrule(lr){6-8}
    \textbf{Features} & one-to-one $\downarrow$ & many-to-one $\downarrow$ & Token $\uparrow$ &
    Boundary $\uparrow$ & Cluster $\uparrow$ & Gender $\downarrow$ & Speaker $\downarrow$\\
    \midrule
    \textit{English} \\
    MFCC & 93.7 & 82.0 & 29.0 & 42.4 & 29.9 & 87.6 & 55.9 \\
    cAE & 93.7 & 82.4 & 28.9 & 42.3 & 29.3 & 83.1 & 49.9 \\
    cAE+VTLN & 93.6 & 82.1 & 29.0 & 42.3 & 29.9 & 75.8 & 44.8 \\
    BNF & \textbf{92.0} & \textbf{77.9} & \textbf{29.4} & \textbf{42.9} & \textbf{36.6} & \textbf{67.6} & \textbf{35.5} \\
    \midrule
    \textit{Xitsonga} \\
    MFCC & 102.4 & 89.8 & 19.4 & 43.6 & 24.5 & 87.1 & 43.0 \\
    cAE & 101.8 & 89.7 & 19.5 & 43.2 & 24.5 & 82.5 & 37.6 \\
    cAE+VTLN & 100.7 & 84.7 & 20.1 & 44.5 & 31.0 & 74.7 & 32.7 \\
    BNF & \textbf{96.4} & \textbf{76.9} & \textbf{20.6} & \textbf{44.6} & \textbf{38.8} & \textbf{65.6} & \textbf{27.5} \\
    \bottomrule
  \end{tabular}
\end{table*}

Table~\ref{tab:segmentation} compares MFCCs, \gls{cae} features (with and
without \gls{vtln}) and \gls{bnfs} as input to the system of \citet{Kamper2017}.
It shows that both \gls{vtln} and \gls{bnfs} help on all metrics, with
improvements ranging from small to more substantial and \gls{bnfs} clearly
giving the most benefit. The effects of \gls{vtln} are mostly confined to
reducing both gender and speaker purity of the identified clusters (which is
desirable) while maintaining the performance on other
metrics.\footnote{Perfectly balanced clusters would have a speaker purity of
  8.3\% for English and 4.2\% for Xitsonga, and a gender purity of 50\% for both
  corpora.} This means that the learned representations have become more
invariant to variation in speaker and gender, which is exactly what \gls{vtln}
aims to do. However, this appears to be insufficient to also help other metrics,
aligning with the experiments by \citet{Kamper2017} that indicate that
improvements on the other metrics are hard to obtain.

On the other hand, \gls{bnfs} result in better performance across all metrics.
While some of these improvements are small, they are very consistent across all
metrics.
In particular, we observe a much higher cluster purity and lower word error rates,
  which both indicate that more tokens are correctly identified. Gender and speaker
  purity have decreased further, which means that the BNFs are even more agnostic
to gender and speaker variations than the cAE features with VTLN.
This shows that the \gls{bnfs} are also useful for down-stream tasks in
zero-resource settings. It especially demonstrates that such \gls{bnfs} which
are trained on high-resource languages without seeing any target language speech
at all are a strong alternative to fully unsupervised features for practical
scenarios or could in turn be used to improve unsupervised systems trained
on the target language speech data.

\section{Conclusions}
\glsreset{bnfs}
\glsreset{cae}
\glsreset{utd}

In this work we investigated different representations obtained using data from
the target language alone (i.e., fully unsupervised) and from multilingual
supervised systems trained on labeled data from non-target languages. We found
that the \gls{cae}, a recent neural approach to unsupervised subword modeling,
learns complementary information to the more traditional approach of \gls{vtln}.
This suggests that \gls{vtln} should also be considered by other researchers
using neural approaches. On the other hand, our best results were achieved using
multilingual \gls{bnfs}. Although these results do not completely match the
state-of-the-art features learned from target language data only
\citep{Heck2017,Heck2018}, they still perform well and have the advantage of
only requiring a single model from which features can be immediately extracted
for new target languages. Our \gls{bnfs} showed robust performance across the
8~languages we evaluated without language-specific parameter-tuning. In
addition, it is easy to control the dimensionality of the \gls{bnfs}, unlike in
the nonparametric models of \citet{Heck2017,Heck2018}, and this allowed us to
use them in the downstream task of word segmentation and clustering. We observed
consistent improvements from \gls{bnfs} across all metrics in this downstream
task, and other work demonstrates that these features are also useful for
downstream keyword spotting in settings with very small amounts of labeled
data~\citep{Menon2018}. We also showed that it is theoretically possible to
further improve \gls{bnfs} with language-specific unsupervised training,
and we hope to explore models that can do this more reliably than the \gls{cae}
in the future.

Finally, our qualitative analysis showed that both \gls{cae} features and
\gls{bnfs} tend to vary much less over time than traditional PLPs, supporting
the idea that they are better at capturing phonetic information rather than
small variations in the acoustics. Although this property helps explain the
better performance on intrinsic measures and the segmentation task, it harms
performance for \gls{utd}, where the system seems heavily tuned towards PLPs.
Therefore, our work also points to the need for term discovery systems that are
more robust to different types of input features.

\section*{Acknowledgements}
This research was funded in part by a James S. McDonnell Foundation Scholar
Award.

\bibliographystyle{elsarticle-harv}
\bibliography{references}

\end{document}